\documentclass[12pt,preprint]{aastex}

\shorttitle{Near Infrared Adaptive Optics Observations of T Tauri}
\shortauthors{Furlan et al.}

\begin{document}


\title{Near-Infrared, Adaptive Optics Observations \\
       of the T Tauri Multiple-Star System}


\author{
Elise Furlan\altaffilmark{1}, William J. Forrest\altaffilmark{2}, 
Dan M. Watson\altaffilmark{2}, Keven I. Uchida\altaffilmark{1}, 
Bernhard R. Brandl\altaffilmark{1,}\altaffilmark{3},
Luke D. Keller\altaffilmark{1,}\altaffilmark{4}, 
and Terry L. Herter\altaffilmark{1} 
}

\altaffiltext{1}{Center for Radiophysics and Space Research, Cornell University, Ithaca, 
NY 14853; furlan@astro.cornell.edu}
\altaffiltext{2}{Department of Physics and Astronomy, University of Rochester, Rochester, NY 14627}
\altaffiltext{3}{Sterrewacht Leiden, PO Box 9513, 2300 RA Leiden, The Netherlands}
\altaffiltext{4}{Department of Physics, Ithaca College, Ithaca, NY 14850}


\begin{abstract}
With high-angular-resolution, near-infrared observations of the young stellar 
object T Tauri at the end of 2002, we show that, contrary to previous reports,
none of the three infrared components of T Tau coincide with the compact 
radio source that has apparently been ejected recently from 
the system \citep{Loinard03}.  The compact radio source and one of the  
three infrared objects, T Tau Sb, have distinct paths that depart from 
orbital or uniform motion between 1997 and 2000, perhaps indicating that their
interaction led to the ejection of the radio source. The path that T Tau Sb took 
between 1997 and 2003 may indicate that this star is still bound 
to the presumably more massive southern component, T Tau Sa. The radio source 
is absent from our near-infrared images and must therefore be fainter than K = 10.2 
(if located within 100 mas of T Tau Sb, as the radio data would imply), still consistent 
with an identity as a low-mass star or substellar object.
\end{abstract}


\section{Introduction}

Stars frequently form in small, gravitationally-bound groups of three or 
more \citep{Ghez93, Mathieu94}, usually situated so that each perturbs 
the motion of the others significantly. In such systems the stellar orbits 
develop chaotically. Over time, gravitational interactions among the members  
and with the residue of the disks from which they formed, result in large orbital 
changes and even ejection, especially for the lighter components 
\citep[see, e.g.,][]{Marchal90}. As main-sequence stars are always seen in 
configurations that are quite different from those of young multiple stars, 
direct observation of orbital evolution in young multiple-star systems has 
long been sought.

Thus the report by \citet{Loinard03} was greeted with wide interest: one of 
the components of T Tauri has, in the last few years, undergone a dramatic 
change in orbital velocity, possibly resulting in ejection from the system. 
T Tau has long been the archetype of young solar-analog stars, and more 
recently has become an archetype of young multiple star systems. Since 
the discovery two decades ago that T Tau is not single, but has a companion
0\,{\farcs}7 to the south \citep{Dyck82}, high-resolution infrared observations 
using speckle interferometry or adaptive optics have been used to resolve the 
system into three stars and to measure accurately the motions arising from their 
orbits: the visible, classic T Tau N, and two heavily-extinguished companions, 
T Tau Sa and T Tau Sb, separated by about 0\,{\farcs}1 \citep{Koresko00, 
Kohler00, Duchene02}. Meanwhile, radio astronomers used the VLA at 2 cm 
wavelength to resolve the system also into two compact components, one 
coincident with T Tau N and a brighter one close to T Tau S \citep{Schwartz86}. 
The southern radio component exhibited motion that until recently was consistent 
with orbital motion around T Tau Sa (which does not appear in the radio images) 
and has been identified with T Tau Sb \citep{Johnston03, Loinard03, Smith03}. 
The radio source is unresolved in VLBI images 
\citep[diameter $<$ 0.5 mas;][]{Smith03}, variable in brightness, and 
exhibits strong and variable polarization in its radio emission 
\citep{Smith03, Johnston03}. This is the object that has apparently been ejected.  

Here we present new near-infrared observations which we use to show that the 
southern radio component cannot be the same as the star T Tau Sb.  
We suggest that T Tau has been a quadruple stellar system, with the lowest-mass 
member probably ejected during the past few years.

\section{Observations}

We observed T Tau under good observing conditions on 2002 December 24 with 
the Palomar Observatory Hale 200-inch telescope, the Palomar Adaptive Optics 
(PALAO) high-order image-correction system \citep{Troy00}, and the Palomar 
High-Angular-Resolution Observer (PHARO) near-infrared camera/spectrometer 
\citep{Hayward01}. The brightest visible component of the system, T Tau N 
(V = 9.6), served as the phase reference for the AO system. We also observed 
SAO 76481 as flux calibrator and point-spread-function (PSF) standard immediately 
after taking our images of the T Tau system. We chose the image scale to be 25 
mas per pixel. 

In order to calibrate the plate scale and orientation of the array, we used our 
observations of the multiple systems RW Aur (two components) and UX Tau 
(four components) and compared the measured separations and position angles with 
the corresponding measurements in \citet{White01}. We chose multiple systems 
whose components have separations wide enough that orbital motion should be 
negligible in the time period between the observations by \citet{White01} 
(1996-1997) and ours (2002.98). The plate scale of our images comes out to 
24.7 $\pm$ 0.4 mas per pixel, and the orientation of the detector array is such 
that north is up and east is to the left with $\pm$ 1$\arcdeg$ accuracy. 
Thus position angles inferred from our data have uncertainty  $\pm$ 1$\arcdeg$.

\begin{figure}
\plotone{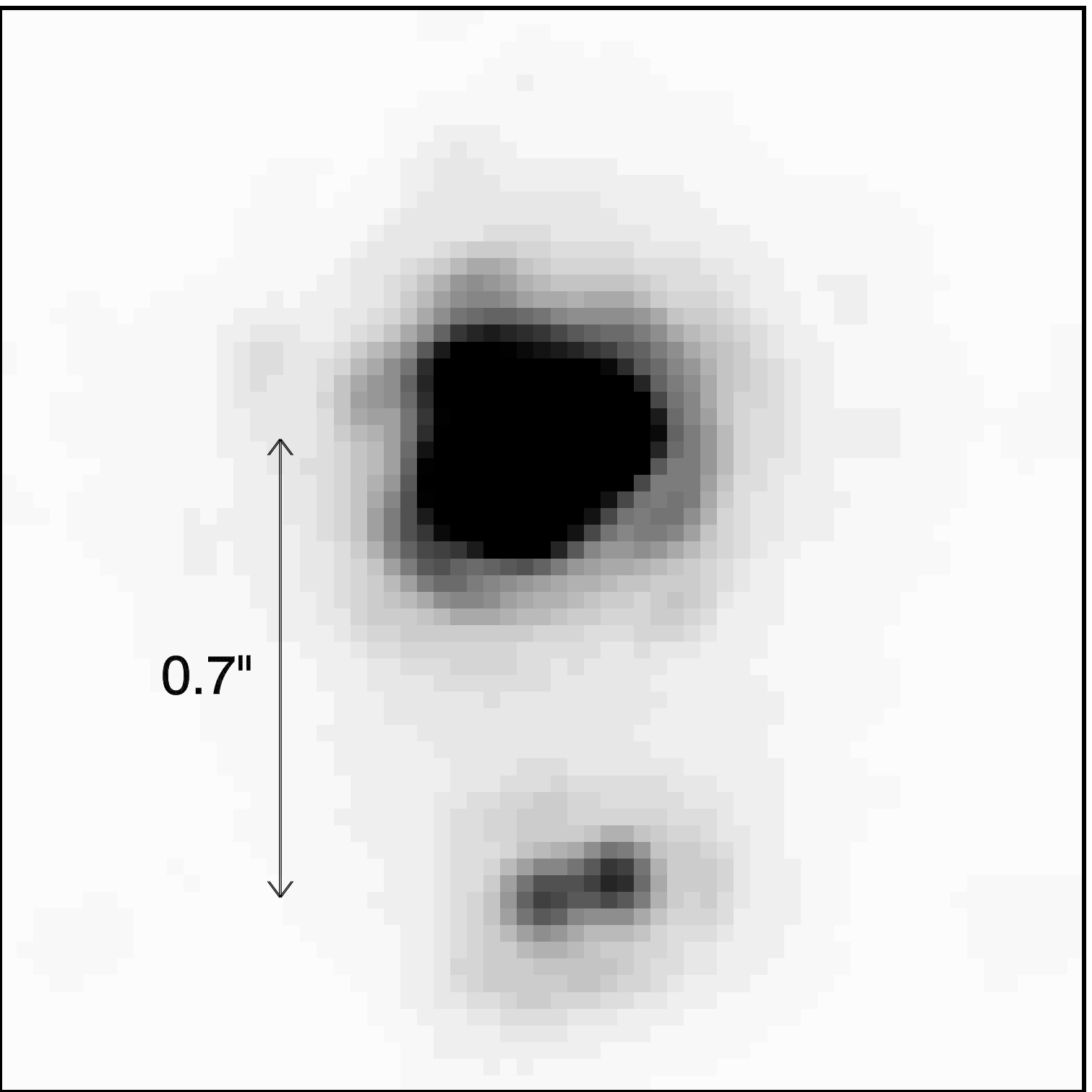}
\caption{Narrow-band 2.26 $\mu$m ($\Delta\lambda$ = 0.06 $\mu$m)
image of the T Tau system, on 24 December 2002: 
T Tau N (top), T Tau Sa (bottom left), and T Tau Sb (bottom right). The image 
is presented in the normal orientation (north up, east left) at a scale of 25 mas 
per pixel, and point sources are 102 mas in diameter (FWHM). Diffraction rings
surround each of the objects; the brightest ring has a radius of about 150 mas.
See Table 1 for 
a list of coordinates and magnitudes. \label{fig1}}
\end{figure}

Figure \ref{fig1} is the image of T Tau which we took in a narrow-band filter 
($\Delta\lambda$ = 0.06 $\mu$m) in the infrared K band centered at 2.26 
$\mu$m. (The components of T Tau are bright enough at near-infrared 
wavelengths that the use of the usual broadband filters is inconvenient, as will 
be seen below). A total of 20 frames, each with exposure time 1.8 seconds, 
were added to create this image. The resulting PSF is close to diffraction-limited, 
102 mas in diameter (FWHM). The limiting 2.26 $\mu$m magnitude for parts of 
the image well separated from the brighter stars is 14.7 (5 $\sigma$ per pixel).  
Within 200 mas (two PSF widths) of the stars T Tau Sa and Sb, diffraction rings 
reduce the limiting magnitude to a range from 10.2 to 11.2. All three of the 
previously-identified stellar components of T Tau appear in Figure 1, well 
resolved from one another. Table \ref{tbl1} is a list of their positions and magnitudes. 
T Tau Sb lies 0\,{\farcs}107, at position angle 289$\arcdeg$, from T Tau Sa, and 
the magnitudes of the two southern components are about the same, with T Tau Sb 
brighter by a factor of 1.05. Since in late 2000 the corresponding separation and 
position angles amounted to 0\,{\farcs}092 and 267$\arcdeg$, respectively 
\citep{Duchene02}, we probably see orbital motion of T Tau Sb around Sa 
(see also Figure \ref{fig3}).

The positions and fluxes of the three observed infrared components of the T Tau system 
were determined by applying a PSF fit to the data.  Standard IDL procedures were used 
to perform a Gaussian fit to the given PSF, the one of SAO 76481, and to apply it to the 
target stars. The position uncertainty resulting from this method is estimated to be 0.2 
pixels, which corresponds to 5 mas. The flux calibration was carried out by using the 2MASS
K-band flux density of SAO 76481; we estimate the uncertainty of the absolute photometry 
to be $\pm$ 20 \%, and that of the relative photometry to be a few percent.

\begin{figure}
\epsscale{0.5}
\plotone{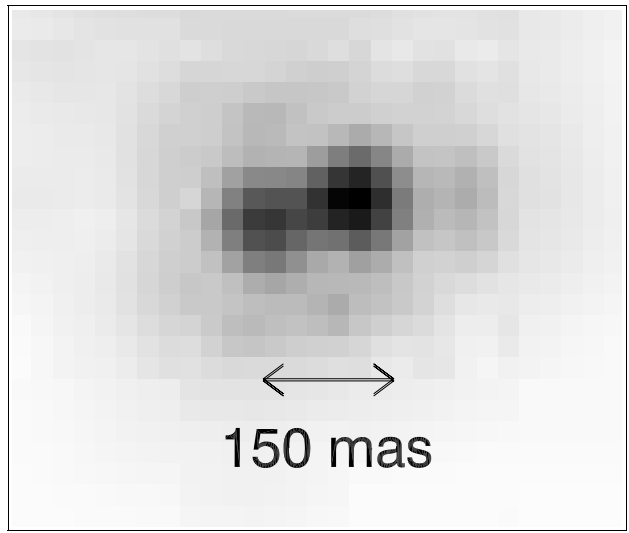}
\caption{2.145 $\mu$m (K$_s$) image of T Tau Sa (left) and Sb (right), 
on 24 December 2002. Orientation and pixel scale are the same as in Figure
\ref{fig1}. \label{fig2}}
\end{figure}

Figure \ref{fig2} is our K$_s$ image ($\lambda$ = 2.145 $\mu$m, 
$\Delta\lambda$ = 0.31 $\mu$m) of T Tau Sa and Sb, taken 5 minutes
before our narrow-band K data. It is also a sum of 20 frames of 1.8 seconds 
exposure time each. Since T Tau N and our calibrator, SAO 76481,
saturated even with this short integration time, we used the image of SAO 76481 observed with
the 2.26 $\mu$m filter to perform PSF-fitting on this image. The position difference
between Sa and Sb is nearly the same in this image and the one taken in the 2.26 $\mu$m 
filter; the offsets listed in Table \ref{tbl1} are averages of the two. No absolute 
flux calibration was possible, but relative photometry was: T Tau Sb is brighter than T 
Tau Sa by the factor 1.36, somewhat higher than we obtained with the narrow-band data. 
In another image taken about 30 minutes later with the K$_s$ filter and an occulting mask
over T Tau N, the flux ratio of T Tau Sa and Sb was the same as in the previous K$_s$
image. The difference between this result and the flux ratio seen in the narrow-band K 
filter (Sb/Sa = 1.05) could be the result of differences in spectral features and degree
of extinction between the two components.

Comparing our narrow-band K data to the results from late 2000 by \citet{Duchene02}, 
T Tau Sa is fainter and T Tau Sb is brighter, each by about a magnitude at K. That T Tau 
Sb can be as bright as T Tau Sa in the K band means that the orbital motion of T Tau Sa 
relative to T Tau N is confused in all pre-1997 measurements because of the presence of 
T Tau Sb.  

\section{Analysis and Conclusions}

\begin{figure}
\epsscale{1.1}
\plotone{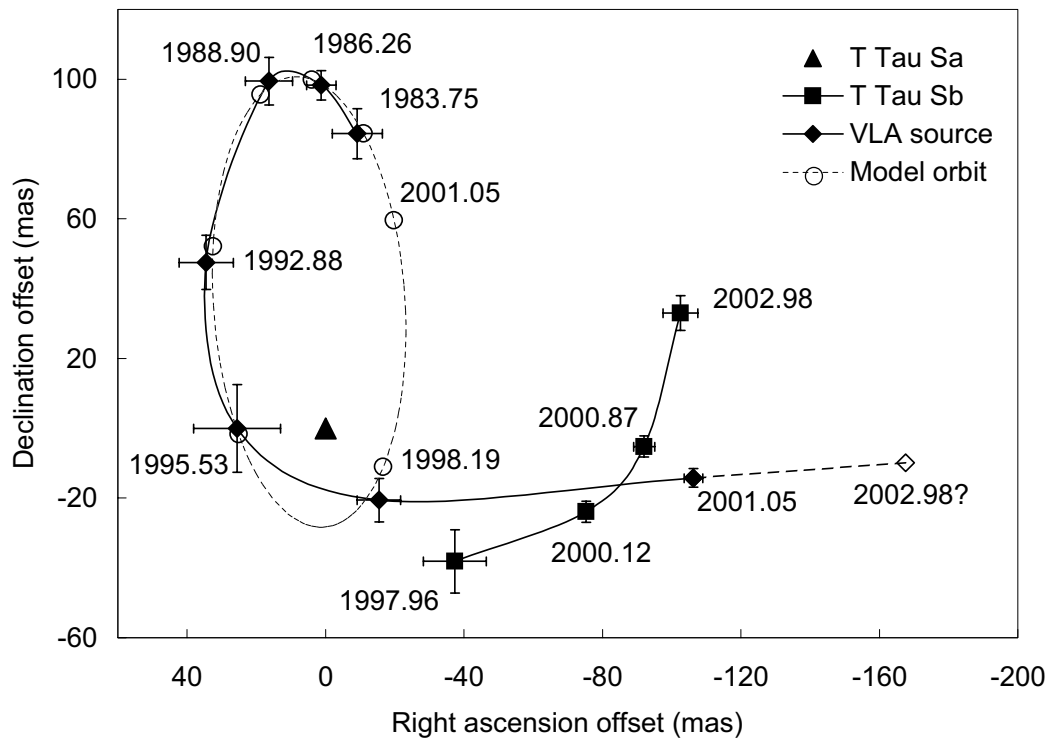}
\caption{Positions of the compact radio source T Tau Sc (solid diamonds) 
and the infrared star T Tau Sb (solid squares), in the rest frame of T Tau Sa (solid triangle), 
labelled by epoch. The positions for T Tau Sc are those reported for the radio source 
by \citet{Loinard03}, shifted east by 4.5 mas for reasons discussed in the text. 
A model orbit, also discussed in the text, appears as a dashed curve, with the points 
corresponding to the VLA radio observation epochs appearing as empty circles. 
The positions of T Tau Sb come from \citet{Koresko00}, \citet{Kohler00}, 
\citet{Duchene02}, and the present work (south to north). Solid curves connecting 
the observations of T Tau Sc and of T Tau Sb are included only to guide the eye. 
A hypothetical epoch-2002.98 position for T Tau Sc, extrapolated linearly from 
the two most recent VLA observations, is plotted as an empty diamond. \label{fig3}}
\end{figure}

Figure \ref{fig3} is a plot of the radio/infrared positions of the southern 
components of T Tau between 1983 and 2003, in a frame of reference 
in which T Tau Sa is at rest. The radio positions in this plot differ very slightly from 
those determined by \citet{Loinard03}. They are based on the radio/infrared registration 
of T Tau N and include the (very small) orbital motion of T Tau Sa with respect to
T Tau N; the uncertainty in the distance between origins of the radio and 
infrared reference frames is about 15 mas (radial, RMS). The relative positions 
of T Tau Sa and Sb were measured directly in the infrared images. 
Our new position for T Tau Sb is quite inconsistent with anything along the 
track of the radio source between 1998 and 2001: it misses by 50-100 mas, 
much larger than the combined uncertainty of about 16 mas.  For example, the linear 
extrapolation of the two most recent radio-source positions, the point indicated by
a empty diamond in Figure \ref{fig3}, lies 78 mas from the 2002.98 observed 
position of the infrared star T Tau Sb. If a single object were to have followed 
the path described by both the radio and infrared positions, its velocity in the 
plane of the sky would have had to undergo major changes twice, once between 
1995 and 1998, and once after 2001. Moreover, in at 
least one of the cases, the single object would have had no obvious partner with 
which to interact. Thus the probability is much greater that the radio and 
infrared observations of the southern components of T Tau detect different 
objects. In anticipation of its detection in future infrared observations, 
we will refer to the radio source as T Tau Sc henceforth. 

Also appearing in Figure \ref{fig3} is a model orbit, resulting from a 
minimum-$\chi^2$ fit to the projected trajectory and transverse velocities 
of T Tau Sc measured at the VLA between 1983 and 1996 
\citep{Loinard03, Johnston03}. In this exercise, we assumed T Tau Sa to be at 
one focus of the orbit. The fit to both trajectory and velocity improved after 
shifting all the VLA positions of T Tau Sc with respect to T Tau Sa by 4.5 mas to 
the east, an amount small compared to the uncertainty of the radio-infrared 
coordinate-system correlation. This small shift is present in the positions plotted 
in Figure \ref{fig3}. The agreement between model and observations is excellent; 
we suggest that the 4.5 mas shift is an improvement upon the previous determination 
of the infrared-radio reference-frame alignment. The parameters of the model orbit 
are as follows: period 19.1 years, semimajor axis 9.1 AU, assuming a distance of
140 pc (leading to a total mass of $2.1M_\odot$ for T Tau Sa and T Tau Sc), axis 
inclination 59 degrees from the line of sight, eccentricity 0.57, argument of periastron 
168.5 degrees, position angle 0.5 degrees for the line of nodes, and periastron epoch 
1997.39.
T Tau Sb is a young M1\,$\pm$\,1 star \citep{Duchene02}, for which the mass is much 
smaller than $2.1M_\odot$; thus we have neglected it in this model.

What is the nature of T Tau Sc? Its small size and large (kilogauss) 
magnetic field, indicated by the VLBI observations \citep{Smith03}, are 
consistent with a star-like object. We did not detect any other object 
besides T Tau N, Sa, and Sb in our near-infrared images. At the position 
extrapolated for T Tau Sc at epoch 2002.98 (see Figure \ref{fig3}), our 
upper limit for the detection of a point source is about 10.2 magnitudes at 
2.26 $\mu$m.
Thus it is either lighter than the eighth-magnitude T Tau Sb, or more 
heavily-extinguished, or both. The velocities of T Tau Sb and Sc within the 
T Tau system are larger than those of T Tau Sa, which, if they have been 
gravitationally bound, indicates that both objects are less massive than T 
Tau Sa. Therefore T Tau Sc is probably an ordinary young star or brown 
dwarf, similar in mass or lighter than T Tau Sb, though quite different from 
this star in its magnetic field strength.
 
The presence of two distinct objects, T Tau Sb and the radio source T Tau Sc, could provide a 
natural explanation for changes in each other's state of motion. As \citet{Loinard03} point out, 
the path of T Tau Sc obeys Kepler's laws for an orbit around T Tau Sa from discovery 
through 1995, but departs strongly thereafter.  The path of the infrared star T Tau Sb with
respect to Sa shows a wide variation in area per unit time; between February 2000 and 
November 2000 the rate of area swept out is more than a factor of two greater than from 
late 1997 to early 2000, or from November 2000 to December 2002. The curvature and 
orientation of this path probably indicate that T Tau Sb has been, and is still, bound to 
T Tau Sa, but longer-term observations will be necessary to demonstrate this. 
In any case, both T Tau Sb and Sc experienced large orbital-motion changes in the 1998-2001 
period, when they were closest together in projection. We suggest that their mutual gravitational 
interaction at this time has resulted in these orbit changes\footnote[1]{If 20 years is a typical 
orbital period over the history of the T Tau S system, then there have been of the order of 
10$^5$ revolutions since formation, during which there could have been only a few 
encounters that resulted in large orbital changes. It would seem that astronomers have 
been extremely fortunate to witness this event.}. 
This would amend the suggestion by \citet{Loinard03} that T Tau Sa has a close companion 
(separation $<$ 2 AU) whose interaction with the radio source T Tau Sc (which they identify 
as T Tau Sb) has resulted in the ejection of the latter; instead, T Tau Sb is a somewhat more 
distant companion that has probably caused the ejection of T Tau Sc, which is a fourth stellar 
or sub-stellar member in the T Tau system. 

\acknowledgments
We are grateful to our telescope operator, Karl Dunscombe, and the instrument
engineers, Rick Burruss and Jeff Hickey, for their outstanding support and help.
This work is supported in part by NASA, through grants for the development 
and operation of the Space Infrared Telescope Facility Infrared Spectrograph: 
Jet Propulsion Laboratory (JPL) contract 960803 to Cornell University, and 
Cornell subcontracts 31419-5714 and 31419-5715 to the University of Rochester.
L. D. K. and T. L. H. acknowledge support from the University Space Research 
Foundation under grant USRA 8500-98-014.
This research has made use of the NASA/ IPAC Infrared Science Archive operated 
by JPL, California Institute of Technology (Caltech), under contract with NASA. It has also 
made use of the SIMBAD and VizieR databases, operated at CDS (Strasbourg, France),
NASA's Astrophysics Data System Abstract Service, and of data products from the Two 
Micron All Sky Survey, which is a joint project of the University of Massachusetts and 
IPAC/Caltech, funded by NASA and the National Science Foundation.


\clearpage


\clearpage

\clearpage

\notetoeditor{Please display this figure as a two-column wide figure in the
final, two-column version of this paper.} 



\clearpage

\begin{deluxetable}{lccc}
\tabletypesize{\normalsize}
\tablecaption{Epoch 2002.98 offsets and magnitudes. \label{tbl1}}
\tablewidth{0pt}
\tablehead{\colhead{Star} & \colhead{$\Delta \alpha$ (mas)}   
& \colhead{$\Delta \delta$ (mas)}  & \colhead{m(2.26 $\mu$m)} }
\startdata
T Tau N    &   $+$20 $\pm$ 5   &  $+$691 $\pm$ 5   &   5.6 \\
T Tau Sa   &         0         &          0        &   8.2 \\
T Tau Sb   &   $-$103 $\pm$ 5  &  $+$33 $\pm$ 5  &   8.2 \\
\enddata

\end{deluxetable}

\end{document}